# MAPPING SYSTEM LEVEL BEHAVIORS WITH ANDROID APIS VIA SYSTEM CALL DEPENDENCE GRAPHS

Bin Zhao

JD.com Silicon Valley R&D Center, Mountain View, California, USA

## ABSTRACT

*Due to Android's open source feature and low barriers to entry for developers, millions of developers and third-party organizations have been attracted into the Android ecosystem. However, over 90 percent of mobile malware are found targeted on Android. Though Android provides multiple security features and layers to protect user data and system resources, there are still some over-privileged applications in Google Play Store or third-party Android app stores at wild. In this paper, we proposed an approach to map system level behavior and Android APIs, based on the observation that system level behaviors cannot be avoided but sensitive Android APIs could be evaded. To the best of our knowledge, our approach provides the first work to map system level behavior and Android APIs through System Call Dependence Graphs. The study also shows that our approach can effectively identify potential permission abusing, with almost negligible performance impact.*

## 1. INTRODUCTION

The first two decades of 21$^{st}$ century has witnessed a huge growth of smartphones, especially during the past several years. In the fourth quarter of 2011, smartphone sales outpaced PC sales for the first time ever, and it was not even close, reported by Canalys [1]. Another milestone for smartphones is that smartphone sales have surpassed those of feature phones in early 2013 for the first time and will continue the prevalence over feature phone sales [2]. Of all smartphone sales, Google's Android accounted for 86.8% in the third quarter of 2018 (3Q18), reported by IDC [3]. In the near seen future, Android will continue to dominate the smartphone market [4].

Due to Android's open source feature and low barriers to entry for developers, millions of developers and third-party organizations have been attracted into the Android ecosystem. With more than 2 million apps and 20 billion downloads in Q3 2018, Google Play now acts as the engine of the application ecosystem [5]. However, it has drawn the attention of attackers and malicious apps. White House estimated that malicious cyber activity cost the U.S. economy between $57 billion and $109 billion in 2016 [6], while mobile malware plays an important role in these attacks. Not surprisingly, over 90% are found targeted on Android [7]. The total global number of malicious Android applications has risen steadily in the last 5 years. In 2013, just over a half million samples were malicious. By 2015 it had risen to just under 2.5 million. For 2017, the number is up to nearly 3.5 million, as reported by SophosLab in 2018 [8].

A malicious application can harm a user's mobile device in various ways. A Trojan malware can steal user's data like contact list and email addresses, hijack user's device resources, and prevent user from performing some actions, etc. Spyware can stealthily collect data regarding user's behaviors and send those data to a remote server. Ransomware can deny access to your mobile phone or data until a ransom is paid, which could be devastating to an individual [9].

Android provides multiple security features to achieve the goal of protecting user data and system resources and providing application isolation. Essentially, those features can be divided into two layers: system and kernel level security, and application level security [10, 11]. System and kernel level security is provided and ensured primarily by the Linux kernel. Specifically, a new inter-process communication (IPC) mechanism is provided by kernel to ensure the security when different applications communicate with each other [11]. These features contribute to the process isolation and application sandboxing. The application level security includes the android permission model, application signing and verification, etc. Android permission model provides "additional finer-grained security features through a 'permission' mechanism that enforces restrictions on specific operations that a particular process can perform, and per-URI permissions for granting ad-hoc access to specific pieces of data" [12]. However, there are still security problems with this finer-grained android permission model. Norton's Joe Keehnast said "very few people actually look through an app's permissions before installing it" [13]. For common users, it could be very unclear about the permissions listed. For example, an application may request for the permission of data connection; however, little is known about what it is using the connection for and when the connection is used. Besides confusion, what is worse is the inherent vulnerabilities with this permission model. Android user-based permission model is per-process based. Android is implemented based on Linux kernel which adopts the principle of privilege-based security scheme. The purpose is to isolate user resources one another. Once a user has a permission (privilege) granted during installation, checks against the user will not be done during running [12]. There is no context-awareness checking for the information flow during run time.

Android APIs and permissions play an important role in Android security system. Permission system is achieved via calling Android API and declaring permissions in the application manifest file, though permission check is done in the system_server process. There are several prior studies on correlating the connection between Android APIs and permissions. For example, a tool called Stowaway built by Adrienne P. Felt *et al.* can determine the set of API calls that an application uses and map those API calls to permissions [14]. PScout, developed by Kathy Au *et al.* [15], could identify Android APIs that could be called by an application and the permissions this API call might need. PScout considers both documented and undocumented APIs. These approaches can find whether an Android application is overprivileged or abusing the permissions, provided that APIs can be effectively tracked. However, Android applications not only use more and more undocumented APIs, but also tend to use their own libraries to evade the Android APIs. A primary reason is that those attacks via calling Android APIs will be detected by tools such as Stowaway or PScout. For example, Z. Zhang *et al.* proposed a transplantation attack which could "spy on users without the Android API auditing being aware of it" [16]. Essentially, this transplantation attack takes out the code from the system_server or mediaserver process and builds its own library to avoid directly calling Android APIs.

Based on this observation, we proposed an approach to map system level behavior and Android APIs. Though Android APIs can be evaded, system level behavior cannot be avoided given that Linux kernel resources are required such as Camera driver or Binder driver. In general, our approach has three steps. Step 1, obtain an application's behavior via a combination of system level tracking and symbolic execution. This step will result in some System Call Dependence Graphs (SCDGs). SCDGs are a clear representation of behaviors for an application. Step 2, co-currently with Step 1, obtain all Android APIs called by this application. Step 3, map SCDGs with Android APIs based on system call entries and timestamps. These steps are done on both benign and malicious applications.

Though this work is not the first to apply system level behavior analysis on Android applications [17-21], this is by far the first attempt to map system level behaviors with Android APIs. Our goal in this paper is not to thoroughly detect malicious applications which evaded Android APIs. Instead, we intend to build a connection and finally a mapping between permissions an application declares and the invoked system level behaviors. This is a basis for further study on detecting possible malicious applications. Overall, this work makes the following contributions:

- We systematically employ system level behavior tracking on Android applications. Their behaviors are dynamically represented by SCDGs.

- To the best of our knowledge, this is the first attempt to map system level behavior of Android application with Android APIs.

- The mapping between system calls and Android APIs can be used to detect malicious applications which try to evade Android APIs to conduct malicious actions.

The rest of this paper is organized as follows. We first shed light on some related work in Section 2. An introduction to Android system and Android applications are presented in Section 3, followed by the current issues associated with Android applications in Section 4. Problem statement is also presented in this section. Our approach of mapping system level behaviors with Android APIs is proposed in Section 5. Section 6 presents a case study to evaluate this approach. Current limitations and some future work are discussed in Section 7. Finally, we conclude this paper in Section 8.

## 2. RELATED WORK

Many efforts have been employed to address the security of Android applications. In this paper, we shed light on some work that our study stands upon.

K. Z. Chen *et al.* proposed an approach of contextual policy enforcement to check the interactions between an application and the Android event system [38]. It could detect sensitive operations being performed without the user's awareness. They proposed a new abstraction of Permission Event Graph (PEG) to detect malicious behaviors. This contextual policy is an important complementary to the current Android system. It can detect some malicious behaviors in Android applications. However, there are several issues with this approach. Precision and efficiency still need to be improved. There is a scalability issue in identifying applications PEGs in another context [38].

A. Reina *et al*. proposed a tool CopperDroid of system call-centric analysis to reconstruct Android malware behaviors [17]. CopperDroid tracks system calls to characterize low-level and high-level Android-specific behaviors. CopperDroid also considers the path coverage issue. However, CopperDroid is dedicated to reconstructing application behaviors including malware, not identifying malicious Android applications. K. Singh proposed context-sensitive permission model MobileIFC to for hybrid mobile applications [39]. The model "allows applications accessing sensitive user data while preventing them from leaking such data to external entities" [39]. MobileIFC can also achieve the information flow control over user contents. Although it is not intended to identify malicious Android applications, it provides a framework for the context-awareness permission model.

H. Cai *et al.* presented DriodCat using a diverse set of dynamic features based on method calls and ICC Intents. Those features were obtained from a dynamic characterization study that revealed behavioral differences between benign and malicious apps in terms of method calls and

ICCs [41]. DroidCat achieved relatively high accuracy with acceptable performance overhead. However, this proposal did not discuss the execution paths that could lead to some potential malicious behaviors. The code coverage issue also could undermine the approach in terms of capturing app behaviors. The defined "diverse" set of features could also have to be consistently maintained and updated regularly.

## 3. BACKGROUND: THE ANDROID SYSTEM

Android is an open-source platform implemented primarily based on Linux kernel, and designed mainly for mobile devices. It usually consists of an operating system based on Linux, middleware, application framework, and some essential applications [22], as shown in Fig.1. Linux kernel lies in the lowest layer in the Android system architecture. Besides some traditional features like memory management, security model, network stack and process management that the original Linux kernel supports, power management and some specific mobile phone related drivers are added into this Linux kernel. Those drivers include binder (IPC) driver, USB gadget driver, and Low Memory Killer, etc [23]. Android middleware includes native libraries and Android runtime. Native libraries are used to support screen display, multi-media services, and web browser, etc. Android runtime contains Dalvik virtual machine (VM) and core Java application libraries [22]. Every application runs in its own Dalvik VM. Unlike conventional Java VMs, which are stack machines, Dalvik is register-based VM [22, 24]. In Android, applications are usually written in Java and compiled to Java bytecode (.class files). A dx tool then convert the .class files into *.dex* (Dalvik EXecutable) format. The reason converting Java code into *.dex* format is that optimization of code is required due to limited resources, such as memory and processor speed [22, 24]. Android application framework is essentially a built-in toolkit and APIs providing services and interfaces to develop applications. It includes Activity Manager, Package Manager, Windows Manager, Telephony Manager, and Content Providers, etc. All of them can simplify the reuse of components.

### 3.1. Android Applications

Android applications lie in the top layer of Android system. They are basically Android packages (APK), archive files containing all the content of apps [25]. In Android, each app runs in its own Linux process. Android app comprises four types of app components: activities, services, content providers, and broadcast receivers [25]. An activity is a single screen that users can interact with, such as an activity of composing an email in an email app. Unlike activities, a service has no user interface; rather, it runs in the background, such as playing music while a user composing text messages. A content provider manages the data among applications. Different content providers manage different types of data, such as contact list, images, and videos. Any app with the proper permissions can interact with the corresponding content providers [22, 25]. A broadcast receiver is used to respond to system-wide broadcast announcements, such as battery is too low [25]. Every app should have an AndroidManifest.xml file [26]. The manifest commonly includes the following information: package name, components of the application, permissions, specific libraries the package need linked to, etc. [22, 26]. Permissions are one of the most important declarations in manifest to fulfill the principle of least privilege and privilege separation. As aforementioned, a permission is a mechanism restricting an application's access to a part of the code or to data on the device. An Android app usually has to explicitly declare the permissions to access the protected part, such as GPS, Wi-Fi, Bluetooth service, broadcasting service, telephony service, camera, SMS/MMS messages, contacts, calendar, and network/data connections. Those access are usually implemented via Android APIs. Permissions requested by an app are granted at app install time. The package installer first retrieves permissions from AndroidManifest.xml and check against the signatures declaring those permissions. If needed, users are then prompted to either allow or deny all permissions the installation package requests. If all granted, the app will be installed successfully. During running time, no checks with the user are done [12].

### 3.2. Android APIs

A framework API is provided on the Android platform that enables applications to interact with the underlying Android system, including resources and services. According to Android [27], the framework API consists of several sets of APIs: a set of packages and classes (e.g., android.bluetooth package and its BluetoothHeadset class), a set of XML elements and attributes for declaring a manifest file, a set of XML elements and attributes for declaring and accessing resources, a set of intents, and a set of permissions that Android applications can request (e.g., READ_EXTERNAL_STORAGE), as well as permission enforcements included in the Android system.

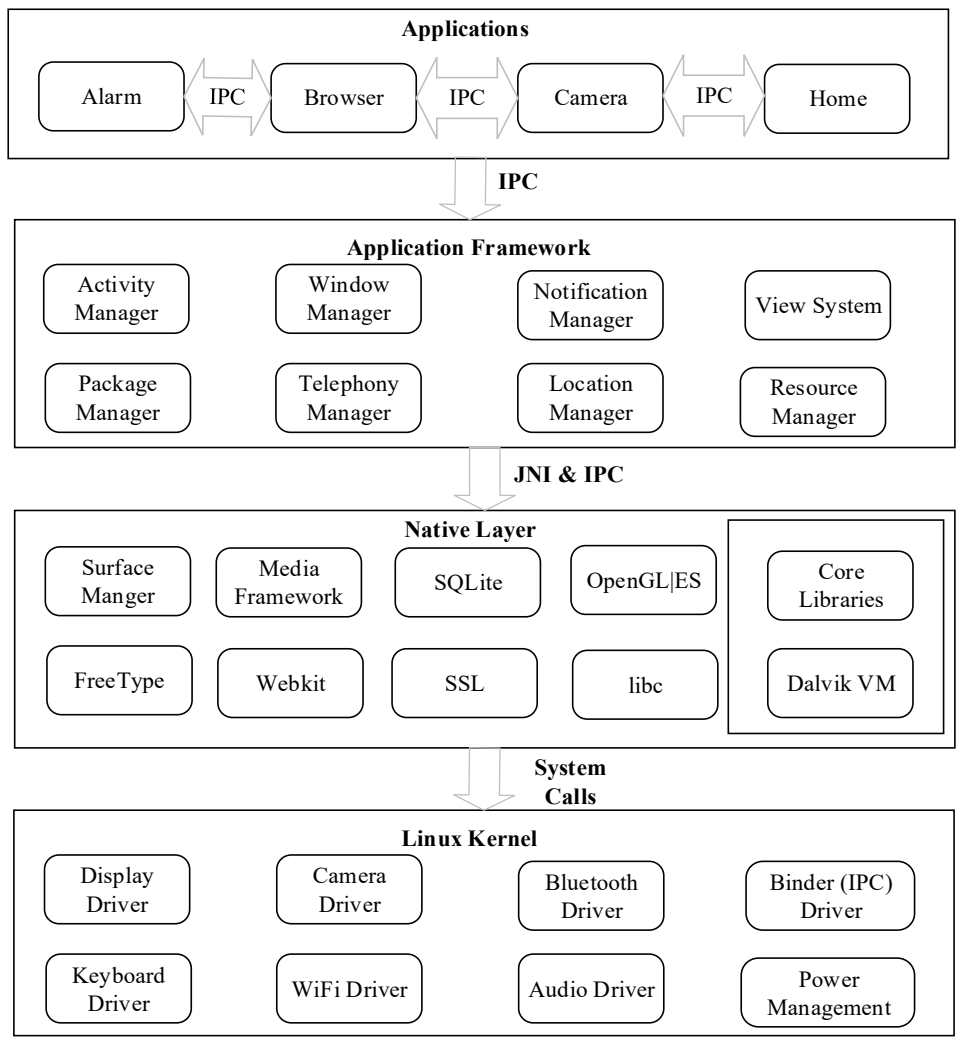

Figure 1. Architecture of Android System

Android applications and system services are all implemented in Java [15]. They access to the lower level libraries via the framework API. The Android application framework implements and provides these Android APIs. They are responsible for the Core system services (e.g., Activity manager) and hardware services (e.g., Location manager). Fig. 2 shows the interaction between an example application process and the system_server process. For example, if an application X wants to use the GPS service. The activity in application X will first call the framework API implemented in framework.jar. In this case, Context.getSystemService (Context.LOCATION-_SERVICE) is called to retrieve a link to android.location.LocationManager for controlling location updates. The Location Manager then get the reference to the Service Manager from the binder. The Location Manager then looks up "location" service in the Service Manager. The Service Manager register as binder context manager, and the references from the binder are obtained. The binder will invoke the LocationManagerService in the system_server process. Permission check is done in the system_server process. If permission of ACCESS_COARSE_-LOCATION or ACCESS_FINE_LOCATION is declared in the android.Manifest.permission, the SystemServer will register "location" service via the Service Manager. The Service Manager finally get the references of the LocationManagerService.

The APIs that an Android application uses can be roughly divided into two categories: documented APIs and undocumented APIs. Documented Android APIs are released on Android platform by Google Android Developers. These documented Android APIs include Android specific libraries (usually start with android, com.android or dalvik), Java compatibility libraries (start with java or javax), and third party libraries (start with org, primarily Apache libraries). Undocumented APIs are also used by Android application developers since Android is open-source. Developers may use Java reflection or implement their own source code to obtain references to any Java method.

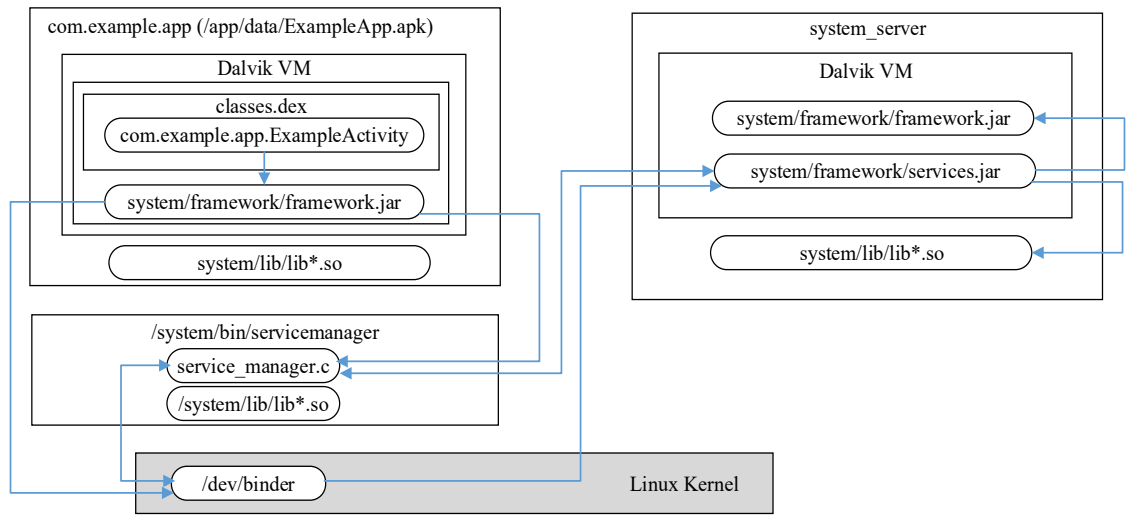

Figure 2. An Example of application calling APIs and binder in Android system. The upper left is an example application process and the upper right is the system_server process.

As mentioned in introduction, to access sensitive resources, permissions should be granted for applications. However, this should work together with Android APIs. As shown in Fig. 2, the permission check occurs in the system_server process. Kathy Au *et al.* found that there is a mapping between Android API calls and permissions [15]. They identified every Android API that could be called and the permissions this API call might need.

## 4. ISSUES WITH ANDROID API AUDITING

In this section, we first do a review of existing techniques for detecting malicious applications, including Android API auditing. We then focus on the issues with Android API auditing. Finally, we present the problem statement of this paper.

### 4.1. Review of Prior Techniques in Detecting Malicious Android Applications

To address permission abusing attacks in Android systems, many works have been presented. Generally, those works can be divided into two primary classes: (C1), static analysis of decompiled applications [17]; and (C2), dynamic analysis to monitor runtime behavior of applications [17, 28]. Static analysis such as DroidRanger [29] and RiskRanker [30] has several strengths. It is relatively easier to decompile Java in Android. Many useful information can be easily obtained in the application's manifest file. In addition, static analysis can properly address the issue of code coverage. However, static analysis is not resilient to obfuscation and compiler optimization techniques. It has a weakness to handle the dynamic code and native code in Android. Sometimes runtime actions cannot be determined unless executed.

Therefore, dynamic analysis such as TaintDroid [28] and DroidScope [31] are aiming to address those weaknesses. They either use instrumentation (TaintDroid) or virtual machine monitor (DroidScope) techniques to conduct the runtime analysis [17]. This technique is resilient to obfuscation and polymorphism. It also properly addresses the runtime action issue. However, code coverage is a big challenge in dynamic analysis. Overhead and scalability can also become a bottleneck for some approaches.

### 4.2. Issues with Current Android API Auditing

Many prior approaches use Android API auditing to detect malicious applications regardless of static or dynamic analysis [14, 15]. However, Android API auditing can be significantly undermined by some existing attacks, e.g. the transplantation attacks. An essential reason is that once an application evades (intentionally or unintentionally) Android APIs whether by native libraries or developers' own implementation, Android API auditing will usually not work effectively. Some malicious attacks like in [16] could even evade the binder.

### 4.3. Problem Statement

Inspired by transplantation attack on Android platform and limitations of existing techniques, we present the problem statement of this paper. First, how to obtain the privileges (permissions) an application actually used? Second, how to represent behavior so that these behaviors can correctly reflect the applications even when these applications evade some Android APIs. This representation should also reflect the functionalities of those applications. Third, how to correlate the permissions of an application with the obtained behaviors? Fourth, can the correlation and representation be used to detect potential permission abuse or privilege escalation?

## 5. DESIGN AND IMPLEMENTATION

### 5.1. Approach Rationale

To address the issues, we propose the system-level behavior analysis on Android applications. There are several reasons that we introduce system-level behavior analysis to identify malicious behaviors. First, system calls are the only interface between Android OS and an application, providing the only way for an application to access the OS services. Second, almost every attack goal is bundled with Android OS resources. Hence, for malicious Android applications, it is usually not possible for them to conduct sensitive actions without triggering system calls, even if they use obfuscation or polymorphism techniques [32, 34]. Third, privilege escalation is commonly achieved through IPC mechanism in Android. System calls used by IPC mechanism in Android can be easily identified and tracked. Fourth, system calls can be practically tracked and analyzed, while giving little overhead to the Android OS, given that Android applications are process-based.

However, from a single system call trace, we know little information about the overall behavior of an application, as system calls are low level reflection about the behavior characteristics of a program. How can we map the low-level system call traces with application level behavior? We need an intermediate representation to connect them together. This is one of the primary reasons for using SCDGs, as SCDGs can appropriately reflect the dependencies between system calls. They are the abstraction of a sequential system call traces. They can clearly describe the interactions among all the private browsing data and disk operations.

### 5.2. Behavior Representation of Android Applications

As aforementioned, an application's behavior can be represented by using a particular graph called SCDG, specifically, a set of disconnected SCDGs. Each SCDG is a graph in which "system calls are denoted as vertices, and dependencies between them are denoted as edges" [32, 33, 40]. A SCDG essentially describes the interaction between a running program and the operating system. This interaction is an essential behavior characteristic of the program [32, 40]. In this paper, SCDG is defined as follows [32, 33, 35].

***Definition 1. System Call Dependence Graph***. *Let p be a running application. Let I be the input to p. f(p, I) is the generated system call traces. f(p, I) can be represented by a set of System Call Dependence Graphs (SCDGs) $\bigcup_{i=0}^{n} G_i : G_i = \langle N, E, F, \alpha, \beta \rangle$, where*

- *N is a set of vertices, $n \in N$ is a system call*
- *E is a set of data dependence edges, $E \subseteq N \times N$*
- *F is the set of functions $\cup f : x_1, x_2, ..., x_n, \rightarrow y$, where each $x_i$ is a return value of system call, y is the dependence derived by $x_i$*
- *α assigns the function f to an argument $a_i \in A$ of a system call*
- *β is a function assigning attributes to node value*

### 5.3. Approach Overview

Fig. 3 shows the architecture of our system. It primarily consists of Dynamic Tracer, SCDG Extractor, IPC Analyzer, Android API and Permission Retriever, and Mapping System.

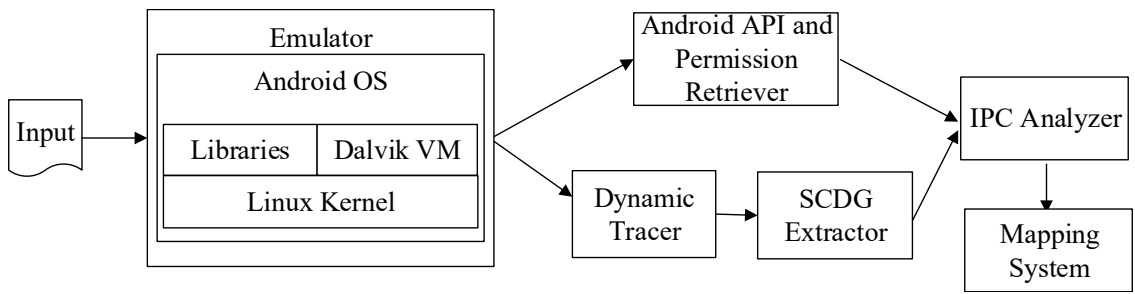

Figure 3. Architecture of mapping system level behavior with Android APIs.

#### 5.3.1 Dynamic Tracer

The dynamic tracer tracks the behaviors of Android applications running on Dalvik VMs in the form of system calls. Android OS is implemented based on Linux kernel. System call invoking is similar even Android OS is running ARM architecture. Hence, we adopt a Dynamic Tracer. A major challenge is that the dynamic tracer needs to trace multiple processes, as permission escalation usually involves several applications as described in Section 4. Noise filtering is required to trim system call traces. *Strace* is used to intercept system calls as it is a built-in tool in Android SDK [36].

### 5.3.2 SCDG Extractor

The SCDG Extractor here takes the trimmed system call traces as the input and aim to generate SCDGs for applications. It primarily explores the dependencies between system calls. API and Permission Retriever. This component has two functions. The permission retriever extracts permission list in the corresponding applications. It retrieves the permissions from AndroidManifest.xml. The output is a list of vectors <*pid, packname, pms$_1$, pms$_2$, ..., pms$_n$*>, where *pid* is the process id of the application, *packname* is the name of the application, and *pms$_n$* is the permission. The API retriever can obtain every Android APIs that are called by an application. This component runs co-currently with dynamic tracer and SCDG extractor.

### 5.3.3 IPC Analyzer

Permission enforcing usually requires the involvement of IPC. Of all the IPC mechanisms, binder acts as the core feature and component. Intents are conveyed via binder. In addition, interactions with the operating system also go through binder [17]. The interactions with the system are primarily through the system call *ioctl()*. The IPC analyzer takes SCDGs as the input and extracts those with IPCs. Finally, based on the arguments, a mapping of caller/callee is presented. A caller is a component in an application that request the permission. A callee is a component in another application that has the permission or is the resource that requested.

### 5.3.4 Mapping System

We use Android's *logcat* as a baseline for mapping system level behavior with Android APIs [37]. All the three use the same time clock and have an accurate timestamp associated with each log, system call entry and API calling.

### 5.4. Implementation

The trace generator is implemented based on *strace* [36, 40]. It can track the system calls and filter off the unnecessary system calls. We implemented the SCDG extractor under *Valgrind*. The SCDG extractor can construct SCDGs. We use *PScout* as the API and permission retriever [15].

## 6. EVALUATION AND CASE STUDY

As mentioned in the introduction, our goal is not to detect malicious applications using this technique. Instead, we intend to build a mapping between the Android APIs and system level behavior. In this section, we present a case study to demonstrate the process of this approach.

### 6.1. Evaluation Environment

The experiments were done on a workstation with a 2.40 GHz Quad-core Intel(R) Xeon(R) CPU and 16GB memory, under Fedora 28, with emulator of Samsung Galaxy S9 on Android 8.0.

### 6.2 A Case Study on Camera Services in Android

We use the transplantation attack in [16] as a case study to show the process of our approach. Overall, it has three steps, obtaining system level behavior represented in SCDGs, retrieving Android APIs, and mapping system level behavior with Android APIs.

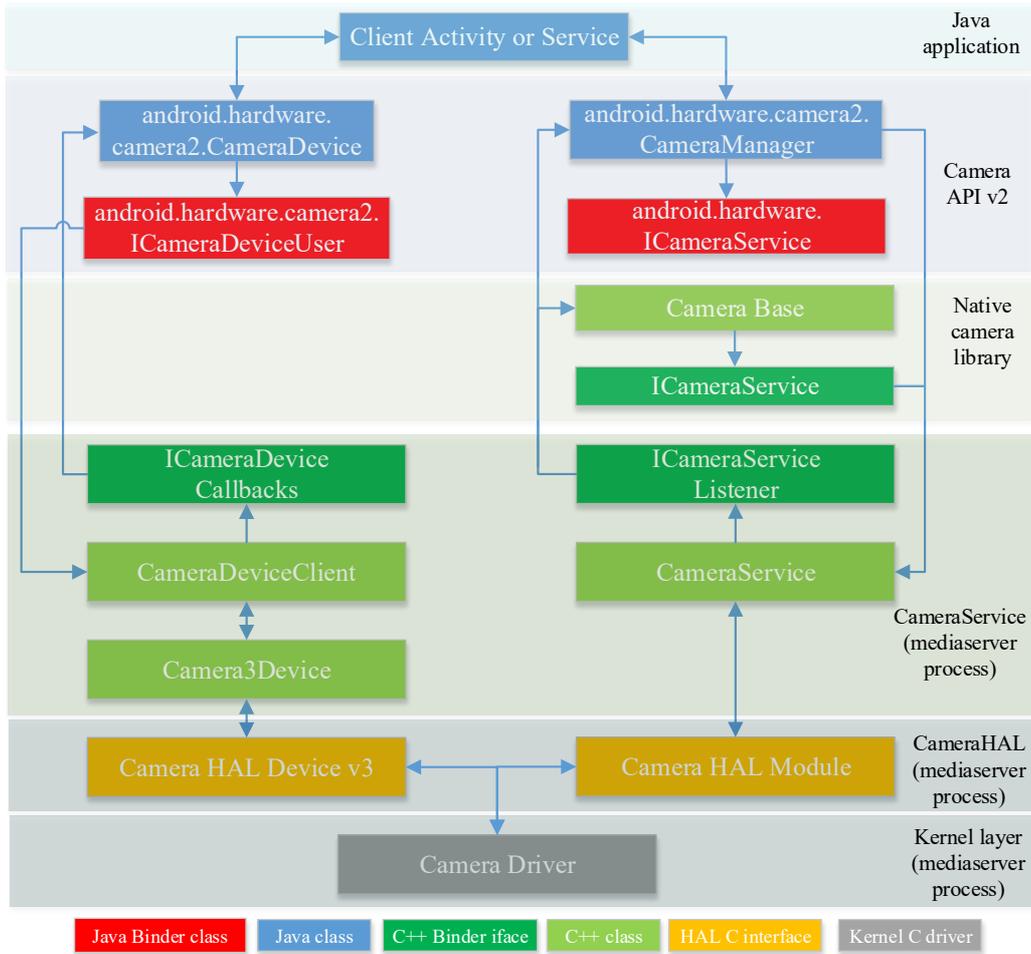

Figure 4. Architecture of camera services in Android system.

**6.2.1 Camera Service in Android System**

Fig. 4 shows the architecture of camera services called by a benign Android application [37]. The camera service is achieved via a native process called mediaserver. The mediaserver contains three layers, native library, HAL and kernel layer (camera driver). A transplantation attack is to transplant the code both in the native library layer and the HAL layer from the mediaserver's address space to the malicious application's address space directly [16]. After transplantation, this application named CameraTest can access to the Camera service and take photos without a user's consent or even awareness.

**6.2.2 Obtaining System Level Behavior**

As every Android application runs in its own process, there is no need to differentiate the system call traces with others. Strace can simply obtain the system call traces for the transplanted application CameraTest process and the mediaserver process. The following Fig. 5 shows part of the SCDGs obtained from CameraTest process.

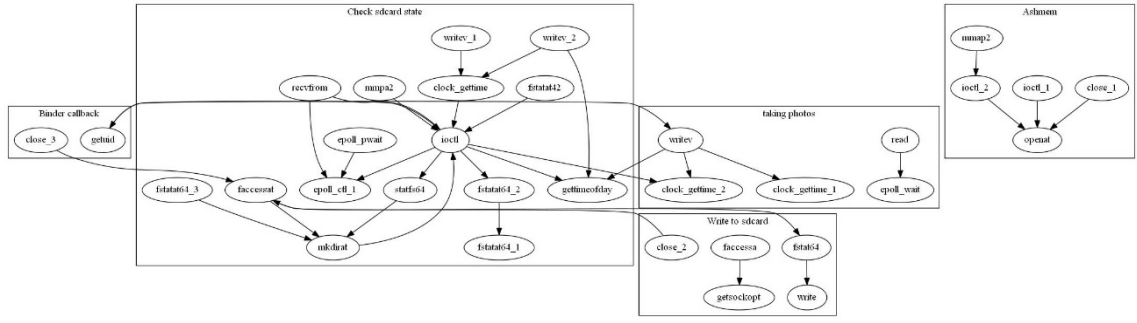

Figure 5. SCDGs obtained from CameraTest process.

### 6.2.3 Mapping System Level Behavior with Android APIs

To map the system level behavior with Android APIs, Android's logcat is introduced in our experiments. Android's logcat collects logs for various applications and portions of the system, where the CameraTest and mediaserver are both included. Based on the timestamps in the system call traces, Android APIs and the logcat, we map the SCDGs to Android APIs. Fig. 6 shows the mapping of SCDGs to Android APIs. The mapping is very clean. For example, the CameraStorage APIs consists just five system calls. It first checks the file status and then uses the binder to implement the IPC interaction. After obtaining the time, it writes to the previously checked folder.

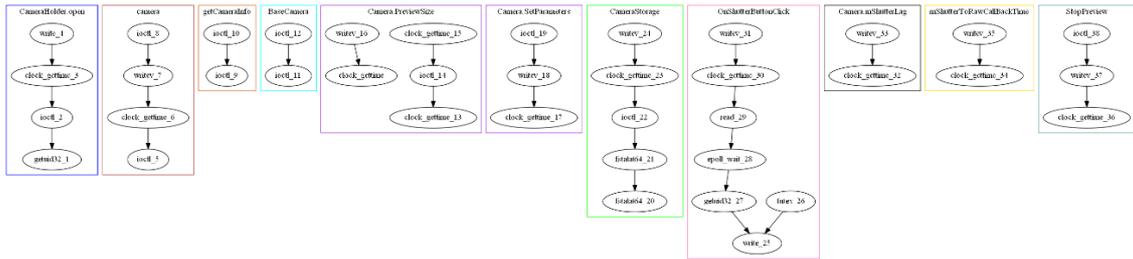

Figure 6. Mapping between SCDGs and Android APIs.

### 6.2.4 Performance Evaluation

We discuss the performance of our system in two metrics: memory usage and startup time.

**Memory Usage**. We use Android's own API to measure the memory usage by calling the *meminfo*. The average memory usage is 1.2% of total memory usage. This memory usage can be considered very small. CPU usage measurement is obtained in Android's Monitoring. The average CPU usage is 0.9%, which is negligible.

**Startup Time**. We calculate the average startup time with 10 tests. The results show that there is almost no difference with or without our system.

Overall, our evaluation on memory usage and startup time demonstrate that the system has very little impact on Android's performance. The overhead brought by the system should not be a concern for users, developers, and app stores.

### 7. DISCUSSION

In Section 5, the dynamic tracer is said to be implemented without the input resolver. In fact, input resolving is a key challenge during the dynamic tracing. An application can result in a set of execution paths due to different inputs, while these execution paths cannot be guaranteed the same during the dynamic tracing. It is very likely that certain malicious actions can only be triggered under specific inputs (i.e., conditional expressions are satisfied, or when a certain command is received). If these specific inputs are not included in the test input space, it is possible that malicious actions can be triggered in a particular execution path. Therefore, an input resolver is still needed to cover more execution paths of applications. Instead of dynamic symbolic execution, behavior stimulation can be used to generate more paths [17].

Most permission policies are enforced by binder driver; however, some permission policies can be enforced by other measures. In this paper, currently, both binder and ashmem mechanisms are considered. The mapping between the SCDGs and Android APIs can be further used to detect if an Android application is suspicious or not. For example, the SCDGs for an Android API are obtained. If there is a suspicious or even malicious Android application evading the using of a sensitive Android API, through system-level behavior analysis, SCDGs of this application can be retrieved. If those SCDGs are isomorphism, it can be claimed that this application implements the functionality of this Android API. This method makes it much harder for evading the Android API auditing tools.

## 8. CONCLUSION

Android provides multi-pronged security features to achieve the goal of protecting user data and system resources and providing application isolation. Android permission mechanism provides a finer-grained security feature to protect user data and system resources. However, permissions can be abused and escalated by malicious applications. We proposed an approach to map system-level behavior with permissions that an application declares. Currently, our study shows that the proposed approach can map system calls with permissions that an application declares, whether benign or malicious applications. It could also identify potential suspicious permission abuse or escalation. However, more experiments are needed to identify malicious applications via investigating the mapping between system calls and Android application permissions.

**AUTHOR**

**Dr. Bin Zhao** graduated from the Cyber Security Lab at the Pennsylvania State University in 2015. He worked for Palo Alto Networks as a Senior Staff Engineer. Currently he is working in JD.com Silicon Valley R&D Center as a Security Architect. His primary interests are network security, software security, black market, mobile data leakage, and network protocol analysis.